\documentstyle[aps,prd]{revtex} \input psfig.tex \begin{document} \vfill
\centerline{\large\bf Perturbative Color Transparency in Electroproduction
Experiments}\par \vskip 0.5cm \centerline{Bijoy Kundu$^a$\footnote{email:
bijoyk@iitk.ac.in}, Jim Samuelsson$^b$\footnote{email: jim@thep.lu.se} Pankaj
Jain$^a$\footnote{email: pkjain@iitk.ac.in} and John. P.
Ralston$^{c}$\footnote{email:ralston@KUHUB.PHSX.UKANS.EDU}} \vskip 0.5cm
\centerline{$^a$Department of Physics, IIT Kanpur, Kanpur-208 016, INDIA}
\centerline{$^b$Department of Theoretical Physics, Lund University, Sweden}
\centerline{$^b$Department of Physics and Astronomy, University of Kansas,}\par
\centerline{Lawrence, KS 66045 USA}

\vskip 1.0cm PACS number(s): 13.40.Fn, 12.38.Bx, 14.20.Dh \vskip 2.0 cm
\centerline{\bf Abstract} We calculate quasi-exclusive scattering of
a virtual photon and a proton or pion in nuclear targets.  This is the
first complete calculation of
``color transparency" and ``nuclear filtering " in  perturbative QCD.  The
calculation includes full integrations over hard interaction kernels and
distribution amplitudes in Feynman -x fractions and transverse spatial
separation space $b$.  Sudakov effects depending on $b$ and the momentum
transfer $Q^2$ are included. Attenuation of the hadronic states propagating
through the medium is calculated using an eikonal Glauber formalism. Nuclear
correlations are included explicitly.  
We find that the color transparency ratio is comparatively
insensitive to
theoretical uncertainties inherent in perturbative formalism,
such as choice of infrared cutoff scales.  
However,  the $Q^2$ dependence of the
transparency ratio is found to depend sensitively on the model of the
distribution amplitude, with endpoint-dominated models failing to be
dominated by short-distance.
Color transparency experiments should provide  
an excellent test of the underlying theoretical assumptions used
in the pQCD calculations.  

\section{Introduction} Exclusive processes are an exciting frontier.
However the applicability of perturbative QCD at the momenta currently
accessible remains controversial. The quark-counting scaling laws of
Brodsky and Farrar tend to agree remarkably well with data. This
apparently indicates that a finite, minimal number of quarks is being
probed.  However, the helicity conservation selection rules of Lepage
and Brodsky tend not to agree with data \cite{GRF,GPL,CZ}. 
Failure
of hadronic helicity
conservation rules out dominance by the short distance
formalism.  Then the agreement of the scaling laws becomes rather mysterious.
Theoretical criticisms focus on calculations found to include regions
where the internal momentum transfers are too small for perturbative
QCD ({\it pQCD}) to reliably apply \cite{Isg,Rad}.  For even the simplest model
calculations, the case of hadronic form factors, it is found that
large contributions come from the
components of quark wave functions involving large quark spatial
separations.  This undermines restriction of the calculation to
short-distance wave functions, which is nevertheless invariably done,
causing problems with the theoretical consistency of the
subject.

In contrast to exclusive processes in free space, it has been claimed
\cite{JB,RP90,PBJ} that the corresponding processes in a nuclear medium will be
theoretically cleaner. Large quark separations will tend
not to propagate in the strongly interacting nuclear medium.  Configurations of
small quark separations, on the other hand, which happen to be the
perturbatively calculable region, will propagate with small attenuation.  This
phenomenon, called nuclear filtering\cite{JB,RP90,PBJ}, is the
complement of the
idea called color transparency \cite{Mueller,brodMuell}. 
In its original rendition,
color transparency \cite{Mueller,brodMuell} was based on having large momentum
transfer $Q^{2}$ select short distance, then free to propagate easily through a
passive nuclear probe. Nuclear filtering uses the nuclear medium in an active
way.

\subsubsection{\it Filtering versus Transparency}
\medskip

The distinction between nuclear filtering and color transparency is
sharpened by considering different kinematic limits. For a given
nucleus (nuclear number $A$), the limit of $Q^2$ going to infinity
should show decreasing attenuation, and ultimately perfect
``transparency'' of a nucleus.  The ``transparency limit'' of
$Q^{2}\rightarrow
\infty$ is unrealistic, however.
The ``filtering limit'' takes $A>>1$ with $Q^{2}$ fixed and large enough to
motivate a pQCD
approach. In this case large $A$ should eliminate many long distance
amplitudes. On this basis, it has been
predicted that 
calculations of exclusive reactions in {\it pQCD} are more reliable in a
large nuclear target than in free-space

\medskip
These remarkable phenomena have some experimental support.
Experimentally one finds that the fixed-angle free space process
$pp'\rightarrow p''p'''$ \cite{Car} shows significant oscillations at
90 degrees as a function of energy. The energy region of oscillations
is not small, but extends over the whole range of high energy
measurements that exist, from $s=6$ GeV$^2$ to $s=40$ GeV$^2$.
The oscillations are not a
small effect, but fill out roughly 50\% of the $1/s^{10}$ behavior, and are
interpreted as coming from interference of long and short distance
amplitudes.  The corresponding process in a nuclear environment
$pA\rightarrow p'p''(A-1)$ shows no oscillations, and obeys the pQCD
scaling power law far better than the free-space data\cite{JB,BT88,PBJ}.
The $A$ dependence, when analyzed at fixed $Q^2$, shows statistically
significant evidence of reduced attenuation\cite{JainRalPRD}.  Note
that 90 degrees is a special point, due to Fermi statistics,
and that experimental study is needed at angles other than 90 degrees.
One cannot conclude from a single experiment that {\it all}
long distance components have been completely filtered away, only that
interference between large and small distance components is different
inside the nucleus, and the long distance components are apparently
reduced compared to in free space.

\medskip

It is interesting, then, that other experiments appear to show the
same phenomena.  Data for the free space energy dependence of
$d\sigma/dt$ for $\gamma p \rightarrow \pi^+  n$ and $\pi p
\rightarrow \pi' p'$ at fixed $90^o$ CM angle shows oscillations quite
like the oscillations seen in $pp \rightarrow p'p''$.   The existence of
this data has not been widely appreciated.  Recent work \cite
{pip} predicts filtering of the oscillation phenomena and two
more cases of the transparency ratio oscillating with energy, 
which may be checked in the 
near future \cite{Gao}.

\subsubsection{Sudakov Effects as Vacuum Filtering}

It has long been known that the transverse separation of quarks in free space
reactions is controlled by effects known as the Sudakov form factor.  The
Sudakov effect is closely related to nuclear filtering.  It is somewhat novel,
but fair, to observe that Sudakov effects are the filtering away of large
transverse separations in the vacuum, enforced by the strict requirements of
exclusive scattering. Li and Sterman \cite{LS} included Sudakov effects for the
pion form factor, arguing that a perturbative treatment become fairly reliable
at momenta of the order of 5 GeV.  As low as 2 GeV, it was found that less than
50 \% of the contribution comes from the soft region.  This countered earlier
calculations, which argued that in free space close to 95 \% of the
contribution
to the form factor comes from the soft region\cite{Isg,Rad}. The situation with
the proton form factor is similar but has a larger theoretical uncertainty
\cite{LS1}. For example, the proper infrared cutoff to be imposed on the
exponent in the Sudakov form factor has been controversial. Jakob et al
\cite{BK} argued that the cutoff used by Li \cite{LS1} does not suppress
all the
end point singularities. By using a different infrared cutoff the magnitude of
the form factor was shown to decrease.  However, an improved and much more
complete calculation \cite{BL} recently incorporated the full two loop
correction to the Sudakov form factor. A very minor modification of the
infrared cutoffs then finds good agreement with data. The remaining
dependence on infrared
cutoff implies that a significant contribution remains from a region of
large distance. 

\subsubsection{\it Calculational Approach}

\medskip

Previous calculations of color transparency phenomena 
have followed several dynamical approaches.  In
one approach, an initial state with size of order $1/Q$ is
postulated, which expands explosively as time evolution
progresses. Different groups use different model dynamics: Farrar, Liu,
Frankfurt and Strikman \cite{flfs} model the process with simple classical
physics. Blaizot \cite {BlaizotHO} and Kopeliovich \cite {KopelHO}
model the time evolution with harmonic oscillator wave functions.
Jennings and Miller \cite {jm} use complete sets in the hadronic basis,
along with experimental matrix elements, to model the time evolution.
Calculations within the different
model dynamics schemes \cite{PBJ} show that the expansion rates depend
strongly on model dynamics and the choice of initial states.

\medskip

In contrast, we follow the pQCD approach. The
impulse approximation for the hard scattering postulates a normal sized
initial state \cite{JR92}.  While the struck state is
full sized, one finds that only the short distance
amplitudes dominate inside the integrations.  The zero-distance 
 wave functions are codified in the distribution amplitude 
formalism, upon which the 
short-distance Sudakov factors are built. 
The perturbative treatment in the impulse approximation
includes ``expansion" or diffusion in the quantum mechanical propagation of
quarks sideways and longitudinally \cite{dourdan,stmalo}.  
We will discuss this in
detail
below.  We will use an eikonal form \cite{GL} consistent with pQCD for the
effects of interaction with the nuclear medium.

\medskip

While made primarily for concept exploration, our calculations include
all effects needed for comparison with data, except for
important fine-tuning of kinematics to match details of experimental
observations. Such details vary from experiment to experiment: in
their absence, we have chosen an idealized kinematic point of zero
momentum transfer to the nucleus. By explicit calculation, this point
has been found to differ with a calculation involving realistic
experimental resolutions to within less
than 10 \%. When experimental kinematics become available we
can include them.  Surprisingly, we find that the main
uncertainty in the nuclear calculation arises from uncertainties in nuclear
medium itself. In particular, uncertainties in the nuclear spectral
functions and correlations are sizable. 
With standard assumptions one can proceed with the
calculation essentially using zero parameters and no model dependence. However,
we find that numerical differences between models of nuclear matter are large
enough to cause significant uncertainties. Indeed, comparison with data shows
that the uncertainties in the nuclear spectral functions and the nuclear
correlations now dominate the theoretical uncertainties, and are larger effects
than, for example, the dependence on the choice of infrared cutoff scale.
This is surprising progress.  

The paper is organized by presenting the kinematic framework for
electroproduction of
pion targets from a nuclear medium in some detail.  This is followed
by the more complicated calculation for nucleon targets. A separate
section gives results and brief comparison with data. 

\medskip

\section{$\gamma^* \pi$ Scattering in a Nuclear Medium}

We briefly review the framework for calculation of hadronic
form factors following Li and Sterman \cite{LS}.
We first consider the case of 
pion.

\medskip

Let $P$ and $P'$ be the incident and outgoing momenta of the
hadrons scattered by the $\gamma^ *$.  From factorization the diagrams are
grouped into 3 kinds: the
power-behaved hard
scattering kernel, the resummed soft and collinear regions responsible
for logarithmic evolution and Sudakov effects, and the
non-perturbative wave functions. In the impulse approximation we
integrate over the respective ``minus'' momenta of partons moving fast
in the proper ``plus'' direction along $P$ or $P'$.  (Our convention is
$k^{\pm}= (k^{0} \pm
k^{3})/\sqrt{2}$).  The conjugate variable $x^{+}$ is the light cone
time variable of the partons, evaluated at zero, setting up the impulse
approximation. The longitudinal $+$ momentum fractions are denoted by the
Feynman variable $x_{i}$ for the $i$-th parton.  We let $b_{ij}$ be the
transverse separation between quarks $i$ and $j$, or
$b$ the corresponding quantity for a single pair of quarks.

\medskip

In the Brodsky-Lepage formalism, 
$Q^{2}\rightarrow
\infty$
is
taken at the first step. The result is that $b$ is set to zero, leaving
convolutions of a hard scattering kernel and distribution amplitudes
that depend only on $x$ and $Q^{2}$. The innovation of Sterman and Li
includes
the Sudakov form factor dependence on $b$ inside the integrations, and
afterwards takes the limit
of large $Q^{2}$.  
Including the $b$ dependence, the pion electromagnetic
form factor can be written as:\begin{eqnarray} F_\pi(Q^2) = \int dx_1
dx_2{d\vec b\over (2\pi)^2} {\cal P}(x_2,\vec b,P',\mu) T_H(x_1,x_2,\vec
b,Q,\mu){\cal P}(x_1,\vec b,P,\mu)\ . \label{fpi} \end{eqnarray}  
Here $${\cal
P}(x,b,P,\mu) = exp(-S)\times \phi(x,1/b) + O(\alpha_s(1/b))\ ,$$ plays the
role
of the hadron wave functions, where $\phi(x,1/b)$ is the meson distribution
amplitude. $T_H(x_1,x_2,\vec
b,Q,\mu)$ is the hard scattering kernel, which after incorporating 
the RG evolution from the 
renormalization scale
$\mu$ to $t$, $t={\rm max}(\sqrt{x_1x_2}Q,1/b)$, is given by\cite{LS},
\begin{eqnarray}
T_{H}(x_1,x_2,\vec b, Q, \mu) &=& exp[-4\int_{\mu}^{t} {d\bar\mu\over\bar\mu}
\gamma_q(
\alpha_s(\bar\mu))] \nonumber \\
&\times& T_H(x_1,x_2,\vec b, Q,t).
\end{eqnarray}
$S$ is
the Sudakov form factor,  
\begin{eqnarray}
S(x_1,x_2,b,Q) = \sum_{i=1}^2\left[s(x_i,b,Q) + s(1-x_i,b,Q)\right]
-4\int_\omega^t{d\bar \mu\over\bar\mu}\gamma_q(\alpha_s(\bar\mu)).
\end{eqnarray}
$\gamma_q(\alpha_s)$ in the above equations is the quark anomalous dimension.
Our symbols are the same as
used by Li and Sterman \cite{LS}, who give explicit formulas for $T_H, s(x_i,b,
Q), \gamma_q$ and so on.
The improved factorization used in \cite{LS} retains
the intrinsic transverse momentum $k_T$ dependence in gluon propagators,
since $k_T$ need not be small compared to $\sqrt{x_1x_2}Q$. In
particular there is a dangerous region if any of the $x_i$
get close to zero. The variable $b$ in Eq. \ref{fpi} is conjugate to $k_{T1} -
k_{T2}$, where $k_{T1}$ and $k_{T2}$ are the transverse momenta of the incident
and outgoing partons. As long as $x_1$ and $x_2$ are not close to their
endpoints,
the dominant scale in the scattering is $\sqrt{x_1x_2}Q$ and the small $b$
region dominates the amplitude. Close to the end points,
$\sqrt{x_1x_2}Q$ may become small. However, the large $b$ region is
strongly damped by the Sudakov form factor. The results for the free space form
factor for the pion using this procedure are given in \cite{LS}. The
authors show
that at $Q^2=5$ GeV$^2$, something like 90\% of the contribution comes from a
region where $\alpha_s/\pi$ is less than 0.7 and hence could be regarded as
perturbative.

\medskip

\subsection{\it The Pion: Nuclear Medium Effects}

The nuclear medium modifies the quark wave function such that \cite{RP90}
\begin{eqnarray} {\cal P}_A(x,b,P,\mu) = f_A(b; B){\cal P}(x,b,P,\mu),
\end{eqnarray} where ${\cal P}_A$ is the wave function inside the medium and
$f_A$ is the nuclear filtering amplitude. The formalism predates Li
and Sterman, and naturally has the same kinematic
dependence (modification of the $b$-space wave function) due to the parallel
between nuclear
filtering and vacuum filtering by Sudakov resummation.  An eikonal
form \cite{GL,durand} is 
appropriate for $f_A$:
\begin{eqnarray} f_A(b; B) = exp\left(-\int_{z}^{\infty} dz' \sigma(b)
\rho(B, z')/2\right) .
\end{eqnarray}
Here $\rho(B, z')$ is the nuclear number density at longitudinal
distance $z'$ and impact parameter $B$ relative to the nuclear center.
We have used the fact that the imaginary part of the
eikonal amplitude for forward scattering is related to the total cross
section, explaining our use of the symbol $\sigma(b)/2$. Finally,
we must include the probability to find a pion at position $B, z$ inside the
nucleus, which we take to be a constant times the probability to find a
nucleon. Putting
together the factors, the transparency ratio $T$ is calculated from
$$ T = {d\sigma_{\rm nuclear}\over A d\sigma_{\rm free\ space}}\ ,
$$
where $A$ is the nuclear number. Some theory groups prefer division 
by a model calculation, 
which introduces a potential model dependence of the definition, 
explaining why we use the original definition of Carroll et al 
\cite{Car}. The nuclear 
cross section is calculated by incoherently adding the contribution
due to individual nucleons. 

\medskip

The inelastic cross section $\sigma$ is known to scale like $b^2$ as 
$b \rightarrow 0 $ in $pQCD$ \cite{Low,gunion}. 
We parametrize $\sigma(b)$ as $k b^2$ and
adjust the value of $k$ to find a reasonable fit to the experimental data.
Introduction of this parameter might be avoided. There is a long
history of relating cross sections to diffractive calculations of the same
kind in pQCD.  For reasons to be explained below, we retain the
parameter here.

\subsection{\it Important Details}

Let us comment on some important details of the calculation.
\medskip

\noindent{\it Nuclear Densities}:
Nucleon number densities were taken from Atomic
Data and Nuclear
Data Tables \cite{Vries}. The pion case uses straight densities as
quoted and then proton case (discussed below) includes nuclear
correlations in the form of an effective density distribution.

\medskip
\noindent{\it Wave Functions}: 
For the x-dependence of wave functions we used the CZ 
and asymptotic distribution amplitudes.
We chose not to complicate the calculation
with
models of the soft $b$-space dependence.  These can be inserted as
necessary: by leaving out such factors, one can more easily see from
inspection the relative effects of Sudakov and nuclear filtering.

\medskip
 \noindent{\it Experimental Momentum Resolution}: In
nuclear calculation we have integrated over 
the transverse impact parameter
B and longitudinal
coordinate $z$ locating the targets in the nucleus.  From translational
invariance, the coherent superposition over the nucleus with net
momentum transfer $\vec K$ includes a phase $ exp ( -i B_{T} K_{T}- i
K_{z}z)$.  The phase is not indicated because we set $\vec K=0$ for the
numerical calculations presented
in Section 3. However, we also repeated the calculations for finite
$\vec K$ to check the dependence on this.  In the region of $ K_{T},
K_{z}$ ranging from -25 MeV to 25 MeV, the results for the Au=197 nucleus
changed by less than $10\%$: specifically decreasing by a maximum of 7.7 \%
for the pion and 8.3 \% for the proton.
For rigorous comparison with
experiments one would want to include the effects of finite $\vec K$
integrated over the same range as experimentally observed.
The treatment of Fermi momentum is of course related, and should be
matched consistently whenever models are used for experimental extraction.

{\it Experimental Subprocess Identification}   The experimental 
extraction of the pion form factor in free space assumes certain 
kinematic criteria are imposed.  A  $t$-channel singularity, and 
consistency with the angular distribution of the spin-zero form 
factor are part of proper ``Rosenbluth separation" extracting the form 
factor \cite {Stoler}.  Experimental  cuts determine whether 
other subprocesses not involving the form factor \cite 
{Carlson} can contribute to the observables, leading to a 
well-defined procedure.  We assume equivalent criteria are applied to 
the experimental study of knocking a pion out of the nucleus, and 
note that this is compatible with the momentum transfer $\vec K$ 
discussed above.   With use of over-determined kinematics such as the 
$BNL$ experiment has demonstrated,  the identification of this 
quasi-elastic subprocess seems quite feasible.

\subsection{\it Expansion}

A controversial element of of color transparency and nuclear
filtering is the topic of ``expansion''.  The term describes the time
evolution of the struck system as it moves through the nucleus.  Some
calculations model this using a hadronic basis assumed to be a complete
set. Experience from nuclear physics calculations are then
brought to the problem.  On the other hand, the foundations are unclear,
because many phenomena
involving quarks-including such basic features as scaling in
inclusive reactions-defy successful description in a hadronic basis.
The transformation between the fundamental quark basis in which
transparency has been predicted, and the hadronic basis of model
calculations, cannot be
explicitly written down. Due to this clash there has been a great deal
of confusion.

\medskip

Perturbative calculations in the quark basis
naturally include time evolution.  The basic element in perturbation
theory is the Feynman propagator, $1/p^{2} +i \epsilon$. The imaginary part
is $i \pi \delta(
p^2-m^2)$.  We transform this partly to coordinate space to see the
time-evolution $U(b, x^+; p^{+})$ 
in light cone time $x^+$ and transverse coordinate $b$,
obtaining:

$$U(b, x^+; p^{+}) \approx \frac{1}{p^+} exp(-i b^{2}p^+/2x^{+} + im^2x^+/2p^+)$$

This has been called ``quantum diffusion", but it represents nothing
more than propagation of a free, relativistic particle from a point source.
Ordinary perturbation theory includes this expansion (and much more)
in the convolution of Green functions over all points linked in the
Feynman diagrams: the series of integrals of
$\Delta_{F}(x^{\mu}-x'^{\mu})\Delta_{F}(x'^{\mu}-x''^{\mu})\ldots$,
somewhat concealed when calculations are done in momentum space.

 \medskip

 The question of expansion, then, is how much physical
 expansion is reproduced by the propagation
 implemented by perturbation theory. In coordinate-space 
the integration regions
include
 light-like displacements much larger than the nuclear
 size and extending over the entire volume of transverse separation
 possible. Whatever the idealizations of factorization arguments, the
 actual calculations include both far off-shell regions from the
 scattering kernel
 associated with short-distance propagation, and nearly on-shell
 regions evolving with proper perturbative quantum mechanical expansion over
 long-distances.  The system interacts with the nucleus over the
 entire process, as the $x, b, B, z$ integrals are totally coupled
 without any separation: Thus, sideways propagation is linked to the
 $z$ propagation.  This fact has been misunderstood, perhaps due to
 attention to the {\it asymptotic} limit in which this same formalism
 has been able to establish that the effects of the
 nucleus are decoupled \cite {RP90}.

 \medskip
 Unfortunately we do not know how to translate the regions of
 integration of the quark variables into the hadronic basis. Given a
 perfect ``rosetta stone'' we could predict exactly what hadronic
 picture applies,  and which details of the hadronic
 spectrum such as the masses and widths of resonances are already
 included, or need to be added.  The situation is exactly like the
 mystery of duality noticed by Bloom and Gilman \cite {bloom} in
deeply-inelastic scattering.
 Twenty years later, there has been little progress in explaining how a
 simple perturbative quark-picture prediction of structure functions
manages to
 interpolate precisely between resonances and successive thresholds of
 daunting complexity in the hadronic basis. On this basis, quark perturbation
 theory definitely reproduce multiparticle continuum hadronic
 states in the time-evolution with some reliability.  However, on the same
basis, pQCD does
 not pretend to reproduce detailed structure at particular momenta
 due to resonances. From
 this we believe that the expansion occuring in our calculations is
 of the nature of a multi-state average when viewed in the hadronic
 basis. It is expansion of one kind, which would not include fine details
of particular resonant
 mixing of states.  It is an open question whether details of the hadronic
 spectrum matter in the problem: the problem is closely related to  
determining the precise kinematic region where pQCD would apply.

\subsection{\it Central Versus Endpoint Wave Function Models}

The calculations we report depend on the models of the distribution 
amplitudes.  For discussion we can classify models as ``endpoint" or 
``centrally" dominated, with typical endpoint models being those of 
$QCD$ sum-rules \cite {CZ,KS}, and typical central models being the 
asymptotic distribution amplitude \cite {GPL}.  Experience 
in free space form-factor calculations teaches us that  endpoint 
models tend to be contaminated by long distance contributions while 
central models tend to be more dominated by short-distance. 
Comparison of experimental data with the pion form factor is fairly 
inconclusive and does not favor either class of models \cite {Stoler}. 
   If one 
allows for some reasonable variations within the classes, for example 
not really believing the normalizations of the QCD sum rule 
predictions, then the ambiguity becomes even worse.  Given this 
situation,  we made calculations using representatives from both 
classes, and without assuming too much is known about the  
normalization of the distribution amplitudes.  We then compare the 
calculations to see what each type of model predicts.

\subsection{\it Discussion}

Elsewhere we have emphasized that quasi-exclusive pion scattering in a
nuclear medium should be very interesting to measure\cite{dourdan,stmalo}.
While the pion's small mass makes large momentum transfers more
difficult, there are reasons to believe that experiments at accessible
momentum transfers should be pursued.

\medskip

First, calculations of meson form factors are comparatively reliable: They are
certainly much better than baryon form factors.  The
pion is uncomplicated compared to the proton, lacking the infamous ``double
-flow" configuration \cite{DM}. The pion also allows fewer covariant wave
functions
that could allow orbital angular momentum to flow. Pion decay
directly measures a short distance wave
function normalization, 
pinning down another variable. Finally, the short-distance 
prediction for the pion electromagnetic form factor is apparently not far
from the data in free space.

\medskip

The upshot is that short distance concepts ``almost work'' for the
pion in free space, and theory is easier.  When one does
not have to rely much on nuclear filtering, it becomes a good
approximation to consider the calculation inside the
nuclear target as a free-space form factor followed by some
propagation.  In that approximation one does not need to know the
form factor, which is argued to cancel out in ratios to free-space
processes.  (Indeed, much of the
theory
literature is locked into the approximation that the form factor
cancels out, because only propagation is calculated.) Under those
ideal conditions, the transparency
ratio as a function of $Q^{2}$ serves its
naive function of
measuring transparency.

\medskip

The general situation cannot be so simple. The instant one
acknowledges that the short distance component inside the nuclear
target is not the same as in free space, then the normalization of
the hard scattering is changed\cite{JainRalPRD}. The effect is not
small when current calculations put 50\% of the amplitude as
``soft'': One cannot then consistently argue that
some universal form factor ``cancels out'' in a naive ratio.
Fortunately one can also study the $A$ dependence at fixed $Q^{2}$ and
convert this uncertainty into productive measurements of
atttenuation. This is discussed in Section 4.

\medskip

On the other hand, the effects from uncertainties in nuclear parameters
are about the same for a pion or proton target. These uncertainties also do
not
cancel out in ratios, do not disappear with increasing $Q^{2}$, and
are not much ameliorated by large $A$. At the moment we simply have to
live with these uncertainties.

\section{$\gamma^*$-Nucleon Scattering in a Nuclear Medium}

The calculation of the proton form factor\cite{LS1} in free-space
is rather more
complicated than for the pion. Underlying the difficulty is that there
are two quark transverse separation scales, each of which must be
controlled by Sudakov and nuclear filtering.

\medskip
The factored
decomposition for the proton form factor derived in \cite{LS1} is
similar to that for the pion but much more complicated.
Our notation
follows Ref. \cite{LS1}  which define the
symbols with lengthy expressions we need not repeat in detail  here.
All wave function 
terms are gathered together in a symbol $\Psi_j$.  The hard scattering
kernels are gathered together in a symbol $\tilde H_j$. There are
altogether 48 separate Feynman diagrams to be summed over, but (as
shown by Li \cite{LS1}) 
symmetries of indices related by permutations reduce the sum
over $j$ to two terms, $\tilde H_1, \tilde H_2$. The permutations
require introduction of symbols $t_{11}, t_{21}$ and $t_{12}$, 
which are notations for
scales that are the larger of $1/b_{i}, x_i Q$. 
Internal quark light cone fractions are denoted by $x$, and
transverse separations by $b_{1}, b_{2}$. There are 4 $x$
integrations, 2 $b$ integrations, and one relative angle integration
$\theta$ to calculate the form factor:

\begin{eqnarray} F_{1}^{p}(Q^{2})&=&\sum_{j=1}^2\frac{4\pi}{27} \int_0^1 (d
x)(d
x')\int_0^{\infty} b_1 d b_1 b_2 d b_2 \int_0^{2\pi} d\theta [f_{N}(cw)]^{2}
\nonumber \\ & &\times {\tilde H}_j(x_{i},x_{i}',b_i,Q,t_{j1},t_{j2})\,
\Psi_{j}(x_{i},x'_i,cw) \nonumber \\ & &\times
\exp\left[-S(x_{i},x_{i}',cw,Q,t_{j1},t_{j2})\right]\; , \label{10}
\end{eqnarray} 
with 
$$ (dx) = dx_1dx_2dx_3 \delta\left(\sum_{i=1}^3 x_i-1\right)\ .$$
Here $f_N$ is the proton normalization constant and $cw$ play the
role of factorization scale, above which QCD corrections give the
perturbative evolution of the wave function $\Psi_j$ in Eq. \ref{10},
and below which QCD corrections are absorbed into the model distribution
amplitude $\phi$.

\medskip

The Sudakov exponent $S$ is given by \begin{eqnarray}
S(x_i,x'_i,cw,Q,t_{j1},t_{j2})&=&\sum_{l=1}^3 s(x_l,cw,Q)+
3\int_{cw}^{t_{j1}}\frac{d\bar{\mu}}{\bar{\mu}}
\gamma_{q}\left(\alpha_s(\bar{\mu})\right) \nonumber \\ & &+\sum_{l=1}^3
s(x'_l,cw,Q)+ 3\int_{cw}^{t_{j2}}\frac{d\bar{\mu}}{\bar{\mu}}
\gamma_{q}\left(\alpha_s(\bar{\mu})\right)\;. \end{eqnarray}

Here $s(x_l,cw,Q)$ 
is called the Sudakov function and its explicit expression is given in
\cite{LS}, while $\gamma_{q}$ is an anomalous
dimension. We have shown the Sudakov form factor to emphasize
explicitly
that the region of large $b$, or regions of $x\rightarrow 0, 
x\rightarrow 1$ is suppressed as $Q \rightarrow \infty$.

\subsection{\it The Proton: Nuclear Medium Effects}

For the proton we use the same procedure for effects of the nuclear medium
as used
for the pion
except that now we use the inverse of the factorization 
scale $cw$
to calculate the attenuation cross section of proton inside the nucleus.
Here $1/w$ is taken to be the
maximum of the three distances in a proton i.e
$1/w=b_{max} = max(b_1,b_2,b_3)$.
We again set $\vec K=0$ for our calculations as used for the pion, 
as the effects of finite $\vec K$ are less than 10 \%.
Overall, then, the calculation of the process in the nuclear target
needs a 9 dimensional integration, which is performed by the Monte Carlo 
method.
Results are presented in Section 4. We now turn to specifying
important details.
\medskip

\subsubsection{\it Calculational Details}
\medskip

{\it Wave Functions and Infrared Cutoffs}: 
As in the pion case, the 
calculations for the proton free-space and nuclear filtered processes 
depend on the model of the distribution amplitudes.  Endpoint models 
tend to greatly exacerbate the problem of long distance 
contributions: at the same time these models tend to be the ones used 
to fit data for the proton form factor.  We will report calculations 
with both the $CZ$ \cite {CZ} and $KS$ \cite {KS} endpoint models, 
and also a central model, and then compare them.

\medskip

There has been some controversy regarding the proper choice of the
infrared cutoff in the Sudakov exponent.  
The factorization scale $cw$  separates the perturbative and the
non-perturbative contributions in the wave function $\cal P$.
The
choice $c=1$ proposed in
\cite{BK} uses the largest distance between the three quarks as the cutoff. It
was found that this choice gave results about 50 \% smaller than the
experimental data for the form factor. One can argue that the result
may be reasonable, if other wave functions (and in
particular, non-zero quark angular momentum) contribute heavily in free space.
On the other hand, in Ref. \cite{BL} it was observed that the largest distance does
not correspond to a physical size of the three quark system. A more appropriate
infrared cutoff might consider a configuration of the quark-diquark
type. The resulting cutoff value $c=1.14$
uses the maximum distance between quark and
diquark.  Remarkably, this small
modification with the KS distribution amplitude 
leads to results in good agreement
with the experiment \cite{BL}.

\medskip

Agreement with experiment can be good, but one can play devil's advocate,
and argue that if various contributions at the 50\% level exist,
perhaps the short distance distribution amplitudes currently in vogue are not
properly normalized, but are effectively renormalized to arrange
agreement with data.  The issue cannot be resolved because
different contributions with different signs may cancel. In our
calculations we chose the infrared cutoff parameter $c=1.14$.

{\it Nuclear Correlations}: Following the procedure of Lee
and Miller \cite {LM}, the effects of short-range correlations were
included
approximately by the replacement  \begin{eqnarray} \rho(B,z')\rightarrow
\rho(B,z')C(|z-z'|), \end{eqnarray} where $\rho(B,z')$ is the nuclear 
density at the longitudinal position $z'$ and impact parameter 
$B$ relative to the nuclear center and $z$ is the longitudinal position 
of the point of hard collision. $C(u)$ is a correlation function
estimated in \cite{MS} to be $C(u) = [g(u)]^{1/2}$ with \begin{eqnarray} g(u) =
\left[1-{h(u)^2\over 4}\right] [1+f(u)]^2 \end{eqnarray} where \begin{eqnarray}
h(u) = 3{j_1(k_Fu)\over k_Fu}\ , \end{eqnarray} \begin{eqnarray} f(u) =
-e^{-\alpha u^2}(1-\beta u^2) \end{eqnarray} with $\alpha=1.1$,  $\beta=0.68$
fm$^{-2}$ and the Fermi momentum $k_F=1.36$ fm$^{-1}$.

\section {Results and Discussions}

Experimental results have been reported as a ``transparency ratio''
between cross sections.  Such ratios depend strongly on what one
chooses in the denominator.
 For the
purpose of reporting our calculations we have divided our cross
sections by $A$ times the cross section from a free proton which was also
the choice adopted by the BNL experiment \cite{Car}.  Thus our
transparency
ratio ``T'' is the ratio of the square of a
form factor with $F_{1}$ Dirac structure for the nuclear target divided
by $F_{1}^2(Q^{2})$.  As mentioned earlier, details of kinematic acceptance
and assumed nuclear spectral distribution are needed for precise
comparison with data. 

\medskip

\subsection{\it The Pion}

\begin{figure} [t,b] \hbox{\hspace{6em}
\hbox{\psfig{figure=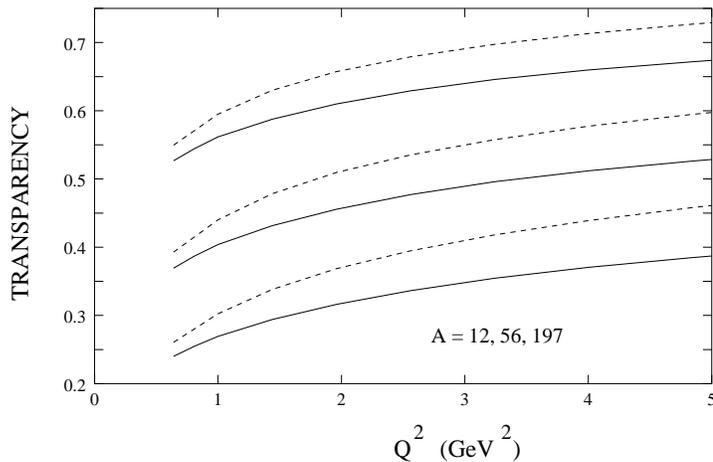,height=6cm}}} \caption{The calculated pion
transparency ratio for different nuclei as a function of $Q^2$. 
The solid and dashed curves use the CZ and asymptotic distribution 
amplitudes respectively and correspond to $A=12$, 56 and 197 from top
to bottom. } 
\label{pionQ} \end{figure}

In Fig. \ref{pionQ} we show the $Q^2$ dependence of the transparency ratio for 
electroproduction of pions using two different distribution 
amplitudes, the $CZ$ and central forms.  The scale of $Q^2$ 
ranging up to 5 GeV$^2$ may benefit from explanation.  At the 
exclusive production point, the relativistic boost factor of a pion 
is given by $\gamma= Q^2/(2 m^2_{\pi}) \sim 25 (Q^2/{\rm GeV}^2)$.  Since 
even a 1 GeV  pion is highly relativistic, we may suppose that the 
perturbative calculations may well apply in the comparatively small 
$Q^2$ regime.  These calculations show a rather striking rise with 
$Q^2$ of the transparency ratio, which should be easily observable 
experimentally.  The fact of the rise does not depend much on the 
distribution amplitude, but the {\it slope} of the rise does: we 
discuss the reasons in the section discussing the proton. 
 For these calculations 
we used $\Lambda_{QCD} = 200$ MeV.  
We adjusted
the value of k so that the predicted results for proton (discussed later) 
are in
agreement with the SLAC data \cite{Mak,Neill}. 
This selects the value of $k$ to be 10. 
This value was determined self-consistently.  It's coincidence with 
the parameter chosen for the pion indicates consistency between the 
two calculations.  Of course, after the integrations are done, 
different regions contribute and the proton tends to have larger 
cross sections than the pion.
 More precise values of $k$ might be obtained
after making a fit to data, or perhaps with more detailed comparison 
with diffractive calculations.

\medskip

\begin{figure} [t,b] \hbox{\hspace{6em}
\hbox{\psfig{figure=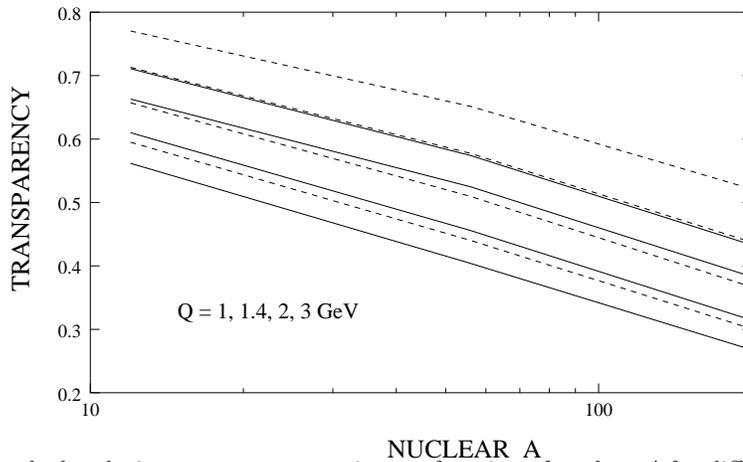,height=6cm}}} \caption{The calculated pion
transparency ratio as a function of nuclear $A$ for different $Q$. 
The solid and dashed curves use the CZ and asymptotic distribution 
amplitudes respectively and correspond to $Q=1,1.4,2$ and 3 GeV from 
bottom to top.
} \label{pionA} \end{figure}

Fig. \ref{pionA} shows the $A$ dependence
of the pion transparency ratio at fixed $Q^2$.  The curvature of the $A$
dependence at fixed
$Q^2$
is a way to extract the effective attenuation cross section independent of the
normalization of the initial state. The normalization of the ratio is
an entirely different affair, which drops out of the extraction of
attenuation cross sections.

\begin{figure} [t,b] \hbox{\hspace{6em}
\hbox{\psfig{figure=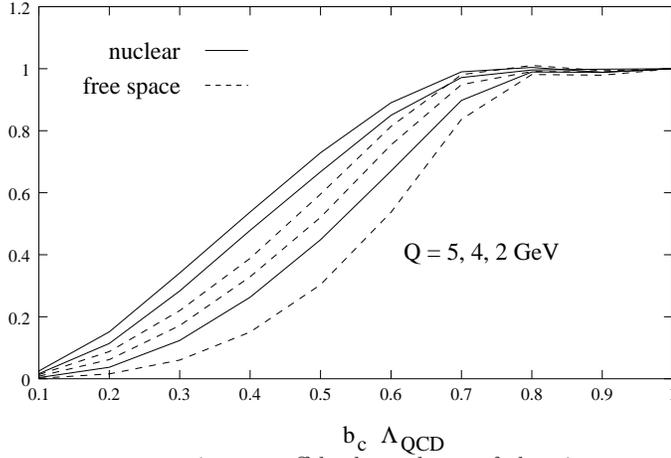,height=6cm}}} \caption{The transverse separation
cut-off $b_c$ dependence of the pion cross section in the nuclear
medium (solid curve) and free space (dashed curve), see the text 
for definitions.
The results have been   
normalized to unity as $b_c$ approaches its maximum value
$1/\Lambda_{QCD}$.  The three different curves in
each case, left to
right, correspond to Q=5, 4 and 2 GeV, respectively.  Calculations are
for $A=197$.
} \label{pionbc}
\end{figure}

\medskip
The pion calculation is
quite transparent, that is, one can easily see the large transverse
separation region being reduced by nuclear filtering.
To quantify this, we introduce a working concept of the {\it cutoff 
dependent cross section}, 
 which we define to be the scattering cross section 
calculated by imposing a cutoff on the quark transverse
separation parameter $b$. 
This terminology should not cause confusion and serves a purpose
for quick visual inspection of $Q^2$ and cut-off dependence.  In Fig.
\ref{pionbc}, we show the cutoff $b_c$ dependence of the pion scattering
cross section ratio. In an ideal short-distance dominated problem, the cut-off
dependence would be absent, and 100\% of the amplitude would occur after
integrating up to $b_c$ of order $1/Q^2$. Cut-off dependence persists, but
compared to free space the nuclear medium
significantly attenuates the large distance contribution. Thus pQCD
is more reliable in the nuclear calculation.

\subsection{\it Extracting the Effective Attenuation $\sigma_{eff}$}

Finally, we have extracted the effective attenuation cross section
$\sigma_{eff} (Q^2)$ , which serve as a litmus test of whether ``color
transparency" has actually been achieved.

\medskip

If one knew for sure that the hard scattering were a short-distance
process, then this procedure would be a complementary test.  However,
when the hard scattering cannot be claimed to ``divide out'' of the
process at realistic $Q^2$, then attenuation becomes central and should
really be
extracted. Following Ref. \cite
{JainRalPRD} we define $\sigma_{eff} (Q^2)$ by fitting the curvature of the
$A$ dependence of the transparency ratio at fixed $Q^2$. In the
process, we let a
($Q^2$ dependent) normalization float. This process eliminates
uncertainties caused by division by a poorly understood free space process:
one can divide by anything fixed, or simply use the cross section in the
nuclear target without division. 

\begin{figure} [t,b] \hbox{\hspace{6em}
\hbox{\psfig{figure=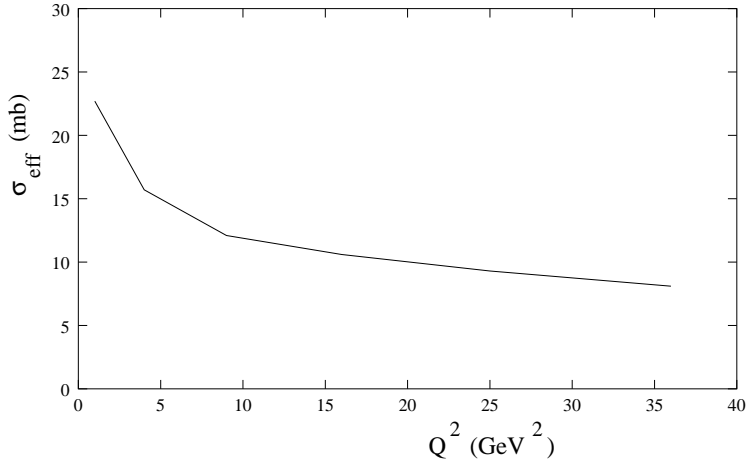,height=6cm}}}
\bigskip
\caption{
Extracted pion effective attenuation cross sections $\sigma_{eff} (Q^2)$ 
as a
function of $Q^2$ exhibit color transparency.  The calculations fit the
curvature of the $A$ dependence
in terms of two parameters $N$ and $\sigma_{\rm eff}$ for each $Q^2$, where
$N$ is the overall normalization of hard scattering in nuclear 
medium and $\sigma$ measures the nuclear attenuation [13].
The decrease of $\sigma_{\rm eff} (Q^2)$ with $Q^2$ indicates that 
pQCD predicts very significant color transparency effect for  
the case of pion. The CZ distribution amplitude has been employed
for these calculations.}
\label{atten}
\end{figure}

\medskip

The results (Fig. \ref{atten}) 
show a significant decrease of $\sigma_{eff} (Q^2)$ with increasing $Q^2$
to values well below the Glauber model attenutation cross section.
The calculations were made in the CZ model; the asymptotic case is
similar.
We found that the normalization ranges from 0.88 to 0.94 as $Q^2$ varies
between 1 GeV$^2$ to 36 GeV$^2$. 
We therefore find that the $\sigma_{eff} (Q^2)$ which
best describes the process is smaller than what would have been
obtained if the normalization were arbitrarily set equal to unity.
Thus the difficulty anticipated on the basis of data some years ago  \cite
{JainRalPRD} has turned out to be a proven feature of the
calculations.

\medskip
The potential scientific value of measuring $\sigma_{eff} (Q^2)$
is very great.  The quantity has the power to rule out conventional nuclear
physics,
independent of rapid $Q^{2}$ dependence, and confirm color
transparency.

\subsection{\it The Proton}
\medskip

In our calculations the parameter $k$ 
in the attenuation cross section
$\sigma=kb^2$
was chosen so as to provide a reasonable fit to the experimental data
\cite{Mak,Neill}. We find that a value of $k=10$ gives a reasonable fit. 
Since the data for $T$ is available only in the region where the
calculated free space form factor is in disagreement with the experimental
result, the values of $k$ obtained by this procedure cannot be taken too
seriously. In fact, parameter $k$ would be best obtained by fitting to the
experimental value of $T$ after it is measured at higher energies. A reasonable
range of $k$ values, which we take to be $k=9$ and $k=10$
has been used in the figures.  We 
note parenthetically that the extrapolation  of the short-distance 
$\sigma =k b^2$ rule into the regime of large $b$ might be modified 
to explore other models.  We did not pursue that here, in order to 
report clearly defined calculations,  perturbatively driven and 
perturbatively consistent.

\medskip

\begin{figure} [t,b] \hbox{\hspace{6em}
\hbox{\psfig{figure=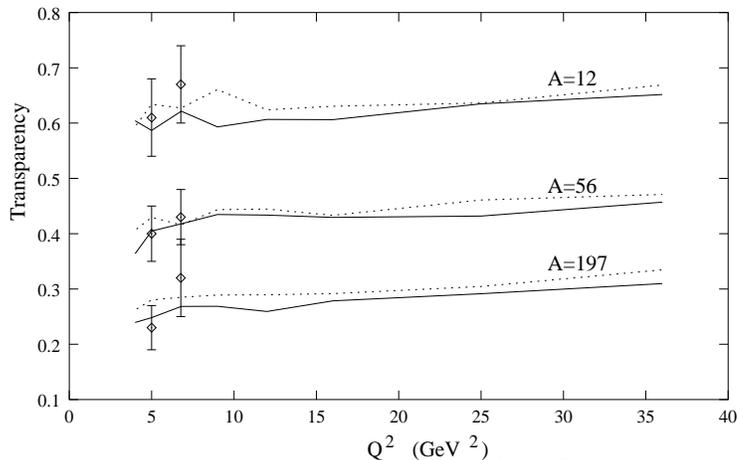,height=6cm}}} \caption{The calculated
transparency ratio for the proton for different nuclei using the
KS endpoint dominated model for distribution amplitude.  The
experimental
points are
taken from Ref. [26,27]. The solid curves are calculated with $k=10$ and the
dashed
curves with $k=9$.} \label{protonQ}
\end{figure}

\begin{figure} [t,b] \hbox{\hspace{6em}
\hbox{\psfig{figure=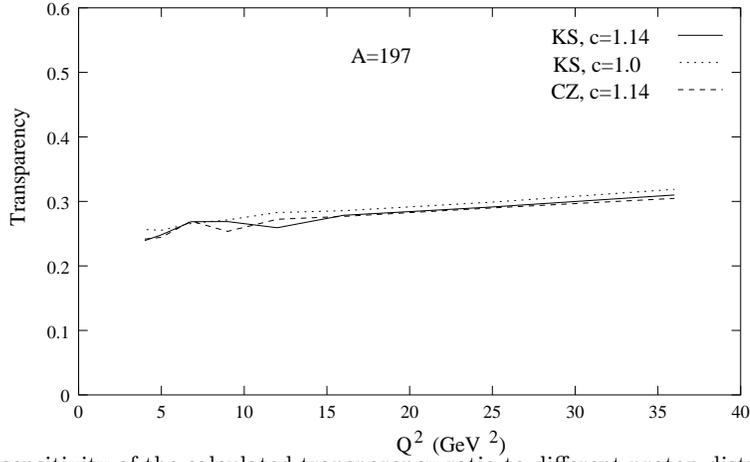,height=6cm}}} \caption{The sensitivity of the
calculated transparency ratio to different proton distribution amplitudes and 
the factorization scale parameter $c$.  
The solid, dotted and dashed curves correspond to
the KS wavefunction with $c=1.14$, KS distribution amplitude with $c=1.0$ and
 the CZ distribution amplitude with $c=1.14$ respectively.
All calculations use $A=197$.}
\label{sensitivity} 
\end{figure}

\begin{figure} [t,b] \hbox{\hspace{6em}
\hbox{\psfig{figure=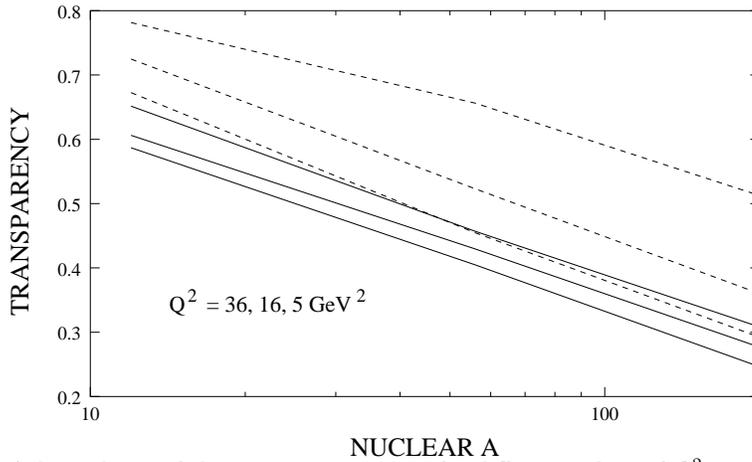,height=6cm}}} \caption{
The $A$ dependence of the transparency ratio for different values of
 $Q^2 = 36, 16, 5$ GeV$^2$ (top to bottom), calculated in the KS endpoint     
 dominated model (solid lines) and a centrally peaked model (dashed lines)
with $k=10$.  The parallelism of the lines in the KS model 
 illustrates next to no change in curvature of the $A$ dependence,
 consistent with little observable signs of short-distance in this
 model.
} \label{protonA} \end{figure}

Results for the $Q^2$ dependence of the proton transparency ratio  
from various models are shown in Figs. (\ref{protonQ}, \ref{sensitivity},
\ref{central}). The nuclear $A$ dependence is shown in Fig. \ref{protonA}.
A standard model
for the distribution amplitude, the KS model, was used to generate
Fig. \ref{protonQ}.  One sees that the calculation with the KS model has a
rather
flat $Q^2$ dependence.  At first this result was surprising, assuming
a short-distance picture and a rapidly increasing function of $Q^2$,
but in retrospect the result appears quite natural. As in 
the earlier discussion of 
the pion, our results depend on the distribution amplitude model, 
which can be categorized into two types. The KS model is
an {\it end-point} dominated distribution amplitude, which is known
 to produce its dominant contributions from long-distance components 
 of the quark wave functions.  For this reason use of the KS wave
 function in the free-space form factor has led to many questions of
 theoretical consistency, which we need not resolve here in this   
 exploratory study.  It is precisely the lack of a substantial
 short-distance contribution which is seen in our calculations to be  
 responsible for the calculated flat dependence on $Q^2$.  Turning to 
 Fig. \ref{sensitivity}, 
which compares the CZ and KS models, both of
which are
 endpoint dominated,  one sees nearly identical flat $Q^2$ behavior. 
 This indicates that the details of the model do not matter so long as
 they are endpoint models.   The figure also shows the dependence on 
 the infrared parameter $c$.  Rather interestingly, a substantial
 dependence on $c$ of form factors in free space drops out in the   
 transparency ratio.

These figures show structure with ``bumps", whose origin
appears to be numerical errors in the Monte Carlo integration.
The error in
the free space and 
nuclear cross section calculation is about 3 \% and 4\% respectively.  
As the accuracy of the integration is increased,
the amplitude of the bumps tends to decrease.  In many cases
  the bumps also are found to randomly shift to different
values of $Q^2$ and hence are consistent with random fluctuations.
Nevertheless the remote possibility that bumps might exist 
in the transparency ratio and be experimentally observable cannot be 
dismissed.

\subsection {\it Long versus Short Distance}

\begin{figure} [t,b] \hbox{\hspace{6em}
\hbox{\psfig{figure=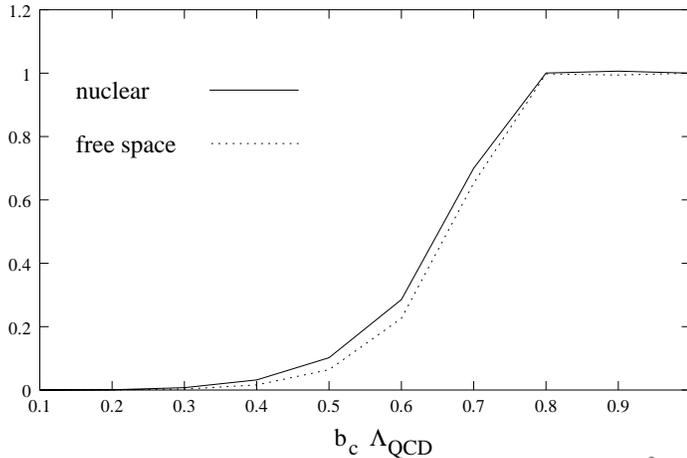,height=6cm}}} \caption{
Dependence on transverse separation cutoff $b_c$ for A=197 and
 $Q^2=36 GeV^2$: see text for details.  Integrations are calculated   
 using maximum separation $b < b_c$ and then normalized to the value
 at the maximum possible $b=1/\Lambda_{QCD}$.
These calculations,
 in the KS endpoint dominated model, show little dominance of short
 distance consistent with the flat transparency dependence with $Q^2$
 of the same model.
} \label{protonbc} \end{figure}

Evidence of dominance by the long distance contributions
  in the endpoint models is shown in Fig \ref{protonbc}, the dependence on a
  transverse separation cutoff $b_c$.  The plot was generated by
  integrating the largest quark separation from zero to an upper limit
  given by $b_c$, and then normalizing the result to the value at the
  maximum $b_c= 1/\Lambda_{QCD}$.   Even with the
filtering effects of
  the nucleus, there is very little saturation and very little
  difference between the nuclear and free-space cases. From Fig.
\ref{protonbc}
we see that the
dominant contribution arises from
a very narrow range of $b$. This appears to be the main reason
why we see greatly reduced filtering effects in the proton
compared to the pion.   Numerical
  experiments showed that by adjusting $k$ arbitrarily, we could force
  substantially more filtering, and a higher proportion of
  short-distance contribution in the end-point models.  However we were
  unable to make such adjustments realistically while maintaining
  reasonable absorption cross sections, and therefore abandoned the
  attempt to make the $KS$ model into a short-distance one. We also note
that we can enhance the short distance contribution in any
model by lowering the value of the infrared cutoff parameter $c$.  Reducing $c$
essentially forces the Sudakov exponent to cut off the integrals at a
smaller value of b.  An enhancement in filtering follows,
which is expected since the integral becomes less dominated
by the region  $b\approx 1/\Lambda_{QCD}$.

\begin{figure} [t,b] \hbox{\hspace{6em}
\hbox{\psfig{figure=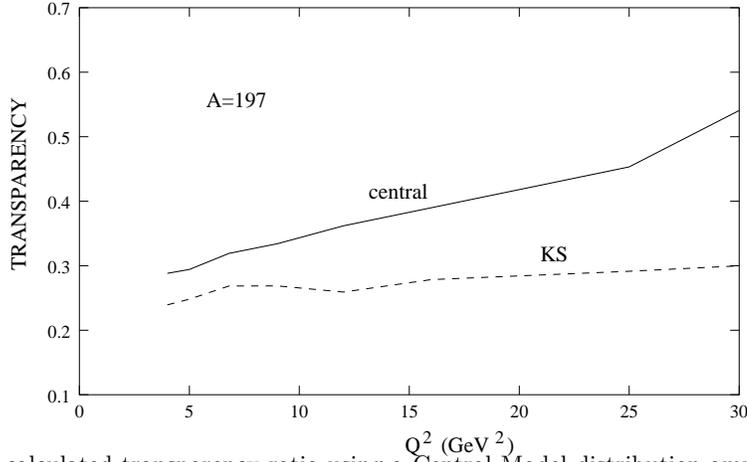,height=6cm}}} \caption{ The 
calculated transparency ratio using a Central Model distribution amplitude 
for $A=197$. The result using KS distribution amplitude is shown for comparison}
\label{central} 
\end{figure}

 Many other wave functions can be conceived: We wholeheartedly believe
 that the wave functions are unknown and that attempts to validate the
 CZ or KS models with data from free-space are rather questionable.
 To explore the effects we took a model which is highly central, a
 distribution amplitude with the structure $$ P(x_i) = x_1 x_2 x_3\ .$$
 In this context we ignored $QCD$ sum-rule lore about the
 normalization of various wavefunctions and simply postulated a
 typical central wave function. The coincidence between this model and
 the asymptotic one derived by Brodsky and Lepage is superficial,
 because any central wave function (such as some Gaussians) would have
 been just as reasonable for the exploration.  We explored the central
 wave function because suppressing the endpoint singularities allows  
 the regions of smaller $b$ to contribute more effectively in the  
 integrations.  The results of the calculation with the central model
 are shown in Fig. \ref{central}.  One sees a rapid rise of the
transparency ratio with increasing $Q^2$ with this model.
 This vindicates the correlation of a rising transparency ratio with
 short distance dominance.

 These results support the following comment on the theoretical
 consistency of the perturbative treatment.  Suppose a rapid rise of  
 the transparency ratio with $Q^2$ were observed: then a central
 model, $A>>1$, and the short-distance assumptions of the calculation,
 would all be internally consistent.  This is in contrast to the
 free-space case, where unfortunately $Q^2$ is not quite large enough
 to draw a similar conclusion, and nuclear filtering is absent.   A
 naive interpretation that central models have been ruled out by
 free-space data also must be set aside as of questionable
 consistency.  Suppose a flat dependence of the transparency ratio
 with $Q^2$ were observed: then one can interpret it as evidence
 favoring the endpoint models, but with the same questionable
 consistency as in free-space. If one is optimistic the nuclear case
 is slightly better in consistency, and then the flat $Q^2$ dependence
 becomes a prediction of the endpoint philosophy. 

 On the last optimistic basis, our calculations indicate that one
 might be able to learn more than previously thought possible 
 about the internal structure of protons from the nuclear target data.
 There is a simple rule: If the transparency ratio rises quickly with
 $Q^2$, then the distribution amplitudes must be centrally peaked.  If
 the transparency ratio rises slowly with $Q^2$, then it is consistent
 with strong end-point contributions.

\medskip

\section {Conclusion} A primary 
uncertainty of color transparency calculations in the models used 
by other groups is the reliability of short-distance assertions of 
the hard scattering.
Without that
assertion, models of hadronic propagation cannot ``divide out'' the
hard scattering.  We do not grant the assertion of idealized
short-distance, and (with all other calculations) find that
contamination from long distance contributions in free-space is ``of
order unity''. However we find
that the long distance components of the amplitudes are considerably
suppressed in the nuclear medium,  compared to the same calculation in
free space. This implies that perturbative QCD is better applied in
the nuclear medium.

\medskip

Using the end point dominated 
model of the distribution amplitude, for the case of the proton we 
find a slow rise in the transparency ratio for energies
that can be probed in the
future at CEBAF and ELFE. The pion, on the other hand, shows a very 
rapid increase. 
It has been known for some years that
the rise with $Q^{2}$ is not particularly
definitive scientific standard \cite{JainRalPRD}.  
Taking into account nuclear filtering, the
slope
measures compensation between two processes: One process tends to
decrease the scattering rate: the
purification of wave functions to short distance.  The other feature
of selecting increased short-distance increases
survival of the struck hadron in propagation. We see this process of
compensation quite clearly in our calculations, and find that it is a
very general feature of pQCD. 
We therefore conclude that 
whether or not a rapid rise with $Q^{2}$ is observed,
the dependence on nuclear number $A$ yields a characteristic curvature from
which effective attenuation cross
sections can be extracted. 

Furthermore we find that 
models of distribution amplitudes
that have wave functions more central in $x$ automatically tend to
have a greater short-distance component, generating a faster  
rise of the transparency ratio with $Q^{2}$. The
slope of $T(Q^{2})$ may serve as a way to probe the fundamental
quark wave functions. The distinction between distribution amplitudes 
which are end point dominated and more centrally dominated is very 
significant for the case of the proton. The end point dominated 
distribution amplitudes give very slow increase in transparency ratio
as a function of $Q^2$ and can be clearly confirmed or ruled in
upcoming experiments at CEBAF and ELFE.

\section * {Acknowledgements} We thank Bernard Pire, Stan Brodsky, Al Mueller
and the organizers of the ELFE Workshop, St. Malo, 1997 for many insights and
discussions in the early stages of this work. We also thank Hsiang-nan Li
for many useful discussions.  Financial support for this work was
provided by
the Board
of Research in Nuclear Sciences (BRNS),
the Crafoord Foundation and
the DOE grant 85ER401214.

\medskip

\section *{References}  \end{document}